\documentclass{emulateapj}

\slugcomment{Draft: \today}

\shorttitle{Cosmic Magnification in the SDSS}
\shortauthors{Scranton et al.}

\begin{document}

\title{Detection of Cosmic Magnification with the Sloan Digital Sky Survey}

\author{
Ryan Scranton\altaffilmark{1}, 
Brice M\'enard\altaffilmark{2}, 
Gordon T. Richards\altaffilmark{3}, 
Robert C. Nichol\altaffilmark{4}, 
Adam D. Myers\altaffilmark{5}, 
Bhuvnesh Jain\altaffilmark{6}, 
Alex Gray\altaffilmark{7}, 
Matthias Bartelmann\altaffilmark{8}, 
Robert J. Brunner\altaffilmark{5}, 
Andrew J. Connolly\altaffilmark{1}, 
James E. Gunn\altaffilmark{3}, 
Ravi K. Sheth\altaffilmark{6}, 
Neta A. Bahcall\altaffilmark{3}, 
John Brinkman\altaffilmark{9}, 
Jon Loveday\altaffilmark{10},
Donald P. Schneider\altaffilmark{11}, 
Aniruddha Thakar\altaffilmark{12}, 
Donald G. York\altaffilmark{13,14}
}
\email{scranton@bruno.phyast.pitt.edu}

\altaffiltext{1}{University of Pittsburgh, Department of Physics and 
  Astronomy, 3941 O'Hara Street, Pittsburgh, PA 15260}
\altaffiltext{2}{Institute for Advanced Study, Einstein Drive, Princeton, 
  NJ 08540}
\altaffiltext{3}{Princeton University Observatory, Princeton, NJ 08544}
\altaffiltext{4}{Institute of Cosmology and Gravitation, Mercantile House, 
  Hampshire Terrace, University of Portsmouth, Portsmouth, P01 2EG, UK}
\altaffiltext{5}{Department of Astronomy, University of Illinois at 
  Urbana-Champaign, 1002 W. Green Street, Urbana, IL 61801}
\altaffiltext{6}{Department of Physics, University of Pennsylvania, 
  Philadelphia, PA 19104}
\altaffiltext{7}{Department of Physics, 5000 Forbes Avenue, Carnegie Mellon 
  University, Pittsburgh, PA 15213}
\altaffiltext{8}{Institut f\"ur Theoretische Astrophysik, 
  Universit\"at Heidelberg, Albert-\"Uberle-Strasse 2, D-69120 Heidelberg, 
  Germany}
\altaffiltext{9}{Apache Point Obs., P.O. Box 59, Sunspot, NM 88349-0059}
\altaffiltext{10}{Sussex Astronomy Centre, University of Sussex, Falmer, 
Brighton BN1 9QJ, UK}
\altaffiltext{11}{Department of Astronomy and Astrophysics,
  Pennsylvania State University, University Park, PA 16802}
\altaffiltext{12}{Department of Physics and Astronomy, The Johns Hopkins 
  University, 3701 San Martin Drive, Baltimore, MD 21218}
\altaffiltext{13}{Astronomy and Astrophysics Department, The University of 
  Chicago, 5640 South Ellis Avenue, Chicago, IL 60637}
\altaffiltext{14}{Enrico Fermi Institute, The University of 
  Chicago, 5640 South Ellis Avenue, Chicago, IL 60637}

\renewcommand{\d}{\mathrm{d}}

\begin{abstract} 

We present an 8$\sigma$ detection of cosmic magnification measured by the 
variation of quasar density due to gravitational lensing by foreground large
scale structure.  To make this measurement we used 3800 square degrees of 
photometric observations from the Sloan Digital Sky Survey (SDSS) containing
$\sim200,000$ quasars and 13 million galaxies.  Our measurement of the 
galaxy-quasar cross-correlation function exhibits the amplitude, angular 
dependence and change in sign as a function of the slope of the observed 
quasar number counts that is expected from magnification bias due to weak 
gravitational lensing.  We show that observational uncertainties (stellar 
contamination, Galactic dust extinction, seeing variations and errors in 
the photometric redshifts) are well controlled and do not significantly 
affect the lensing signal.  By weighting the quasars with the number count 
slope, we combine the cross-correlation of quasars for our full magnitude 
range and detect the lensing signal at $>4\sigma$ in all five SDSS filters.  
Our measurements of cosmic magnification probe scales ranging from 
$60\,h^{-1}\,$kpc to $10\,h^{-1}\,$Mpc and are in good agreement with 
theoretical predictions based on the WMAP concordance cosmology.  As with 
galaxy-galaxy lensing, future measurements of cosmic magnification will 
provide useful constraints on the galaxy-mass power spectrum.
\end{abstract}


\section{Introduction}\label{sec:intro}

We expect the large-scale structure seen in the low redshift Universe
to gravitationally lens background sources, such as high redshift
galaxies and quasars.  This lensing effect causes both a
magnification and a distortion of these distant sources. The systematic
distortion of faint background galaxies by gravitational lensing, the
\emph{cosmic shear}, has now been measured by several groups in the
past few years (\citealt{2000A&A...358...30V, 2000MNRAS.318..625B,
2001ApJ...552L..85R, 2002ApJ...572...55H, 2002A&A...393..369V,
2003AJ....125.1014J, 2003MNRAS.341..100B, 2004astro.ph..4195M}), and
has been found to be in remarkable agreement with theoretical
predictions based on the Cold Dark Matter model. It has also provided
new constraints on cosmological parameters, especially on $\sigma_8,
\Omega_m$ and the shape of the dark matter power spectrum (for a
review, see \citealt{van03,ref03,hoe03b}).  In addition to shear-shear 
correlations, the cross-correlation of foreground galaxies with background 
shear, known as galaxy-galaxy lensing has also been measured 
(\citealt{bra96,del96,gri96,hud98,fis00,wil01,mck01,smi01,hoe03a}).  Recent 
measurements from the Sloan Digital Sky Survey (SDSS;~\citealt{yor00}) have 
enabled accurate constraints on galaxy halo profiles or more generally the 
galaxy-mass correlation (\citealt{sel04,she04}).

In a similar way, the systematic magnification of background sources
near foreground matter over-densities, the \emph{cosmic magnification}, can 
be measured and can provide largely independent constraints on cosmological 
parameters.  Gravitational magnification in the weak limit has two effects: 
First, the flux received from distant sources is increased, resulting in a 
relatively deeper apparent magnitude limited survey.  Second, the solid angle 
is stretched, diluting the surface density of source images on the sky.  
The net result of these competing effects is an induced cross-correlation 
between physically separated populations that depends on how the loss of 
sources due to dilution is balanced by the gain of sources due to flux 
magnification.  Any type of background source can be used to measure 
this effect (galaxies, quasars, supernovae, etc.), but in practice, previous 
investigations have used foreground galaxies and background quasars motivated 
by the large redshift range probed by quasars and general redshift segregation 
between these two populations.  Despite the apparent elegance of this 
solution, lensing-induced quasar-galaxy correlations have been a controversial 
subject for more than a decade.  Numerous teams have attempted to measure this 
effect and have reported detections of changes in the density of background 
quasars in the vicinity of galaxies.  However, as seen by reviewing the 
literature on this type of measurement, the results have been generally 
discrepant with each other as well as in disagreement with the expected 
signal from gravitational lensing.

The first analysis of quasar-galaxy correlations was done by \citet{sel79} 
using a sample of $\sim 400$ quasars and galaxies from the Lick catalog that 
led to a $3.7\sigma$ detection of an quasar excess on $\sim20'$ scales in 
the vicinity of galaxies.  The first measurements aimed at detecting the 
expected lensing signal used radio-selected quasar samples; this method 
yielded numerically larger quasar samples as well as a steeper number count 
relation to enhance the lensing signal.  \citet{FU90.1} correlated bright, 
radio-loud quasars at moderate and high redshifts with galaxies from the Lick 
catalog and found an excess on a 10$'$ scale. \citet{BA93.2} repeated the 
analysis with 56 $z\ge0.75$ optically identified quasars from the 1-Jansky 
catalog and confirmed the previous result. Similar excesses were also found by
cross-correlating the 1-Jansky quasar catalog with IRAS galaxies
(\citealt{BA94.1,BA97.2}) and diffuse X-ray emission (\citealt{BA94.5,coo99}). 
\citet{rod94} found a correlation between optically-selected quasars and 
Zwicky clusters, but with an amplitude that cannot be reproduced by lensing 
of simple mass models. \citet{SE95.2} revisited the previous 1-Jansky/IRAS 
analysis, finding agreement for the intermediate redshift quasars but failing 
to detect any correlation for the high redshift ones.  \citet{WU95.1} repeated 
the cluster cross-correlation using Abell clusters and found no correlation 
with the 1-Jansky sources.  Using wide-field R-band images, \citet{nor99} 
detected a correlation with 1-Jansky quasars on scales greater than 10$'$. 
\citet{wil98} and \citet{nor00} cross-correlated LBQS and 1-Jansky quasars 
with APM galaxies (\citealt{mad90}) and claimed significant over-densities 
on angular scales of the order of one degree.

Using optically selected sources, similarly mixed results have been obtained: 
by correlating UV-excess quasars and APM galaxies in clusters,
\citet{boy88} found a $30\%$ deficit of quasars on scales of 4$'$ around 
galaxies.  \citet{cro99} investigated the lensing explanation but
found the amplitude of the signal to be too high. \citet{wil95}
used variability-selected quasars and found a correlation with
Zwicky clusters that they interpret as induced by lensing. Again, the
amplitude of the correlation largely exceeded expected results.
Associations between red galaxies from the APM catalog and
moderate-redshift quasars were investigated by \citet{ben95a}. They
reported a $\sim 30\%$ excess of quasars within 2$'$ from the
galaxies and a signal consistent with zero on larger scales. 
\citet{fer97} cross-correlated optically selected bright quasars and
galaxies. The amplitude and the redshift dependence of their results
was inconsistent with either the lensing or the dust explanation, and
they suggested that the quasar selection process suffered from
incompleteness.  More recently, \citet{myers03} investigated
correlations between galaxy groups and optically selected quasars from
the 2dF survey. They found a $3\sigma$ anti-correlation within
10$'$, whose amplitude implies that the velocity dispersion of
galaxy groups is of the order of 1000 km s$^{-1}$, i.e. much higher than
expected. In addition, they have investigated the effects of
extinction by dust and found them to be negligible.  Finally, \citet{gaz03} 
measured the cross-correlation between photometric galaxies and spectroscopic 
quasars using only the SDSS EDR (\citealt{sto02}).  In contrast to 
\citet{myers03} they found a positive cross-correlation of 20\% on arcminute
scales -- two results that might be complementary because they probe
quasar samples of different apparent magnitude (see, e.g., \citealt{myers05}),
although the amplitude of both measurements was far in excess of the expected 
lensing signal. Compared to our measurements, the \citet{gaz03} analysis was 
done on earlier reductions of EDR data and an incomplete sample of quasars, 
both of which may have affected the observed signal (see \S\ref{sec:data_sub} 
for more details).  In addition, the foreground and background samples used 
appear to have significant redshift overlap, which can lead to a much stronger 
non-lensing correlation.

Clearly, the scatter in the existing observational results is large,
ranging from significant positive correlations to null and negative 
correlations, as well as a variety of claimed scales for the different 
detections, from half an arcminute to one degree.  In addition, 
quasar-galaxy correlations have been controversial for many years due 
to the large discrepancy between the claimed detections and early theoretical 
estimations (\citealt{sch92, bar95, dol97, san97}).  Indeed, the observed 
amplitude of previous magnification results was typically an order of 
magnitude larger than predicted by theory.  Initially, it was not clear 
whether this problem was due to the observations or the models, as the early 
formalism describing cosmic magnification used several simplifying
assumptions: a linearized magnification and a constant bias for the
galaxy-dark matter relation.  Improvements have recently been made on
the theoretical side, i.e., \citet{men02b} went beyond the linearized
magnification approximation by including non-linear corrections to the
relation between the magnification and density fluctuations, and then
compared their results to numerical simulations. \citet{gui01}
introduced a scale-dependent galaxy bias obtained from measurements of
galaxy auto-correlation functions. \citet{jai03} modeled the complex
behavior of the galaxy bias using the halo-model approach, and
\citet{tak03} added an estimation of the full non--linear
magnification contribution by including the magnification profile of
NFW halos (\citealt{nav97}) in the halo-model formalism. Together, these
works provide an accurate theoretical framework and show that earlier
estimations of the amplitude of the cosmic magnification were
underestimated by 20 to 30 \%.  However, this remains insufficient to reconcile
the expected signal with the observed cross-correlations.  

In this paper, we present the detection of cosmic magnification using
a large, uniform sample of photometrically-selected SDSS galaxies and quasars. 
Contrary to previous results, we find an excess of bright ($g \lesssim 19$) 
quasars around galaxies and a deficit of fainter quasars, matching the 
expected variation with quasar number count slope.  In addition, the amplitude
of the signal and its angular dependence is, for the first time, in agreement 
with theoretical predictions.  Based on a number of data quality and 
uniformity tests, we find that this detection is robust against possible 
sources of systematic error that may have plagued previous measurements and 
represents a genuine detection of magnification bias.  The outline of the 
present paper is as follows: \S\ref{sec:modeling} reviews the basic weak 
lensing models for the expected signal and \S\ref{sec:data} describes the 
galaxy and quasar data sets and cross-correlation estimators.  
\S\ref{sec:results} summarizes the results from magnitude-limited quasar 
samples as well as optimally-weighted measurements using the full quasar 
sample and compares them to the expected signals derived in 
\S\ref{sec:modeling}.  Finally, \S\ref{sec:discussion} discusses the 
possible applications for further measurements using the SDSS and future 
large area surveys.

\section{Modeling Magnification Statistics}\label{sec:modeling}

In this section we briefly describe the formalism of cosmic
magnification and introduce the notation that will be used below.
Let $\mathrm{n_0}(f)\,\d f$ be the number of sources with a flux in
the range $[f,f+\d f]$ and $\mathrm{n}(f)\,\d f$ the corresponding
number of lensed sources undergoing a magnification $\mu$.  We write
the unlensed source counts as
\begin{eqnarray}
\mathrm{n_0}(f)\,\d f &=& a_0\,f^{-s(f)}\,\d f\,,
\end{eqnarray}
where $a_0$ is some normalization factor and $s(f)$ is the power-law slope
as a function of flux $f$.  The magnification effect will enlarge the sky 
solid angle, thus modifying the source density by a factor $1/\mu$, and at 
the same time increase their fluxes by a factor $\mu$. These effects act as 
follows on the number of lensed sources:
\begin{eqnarray}
\mathrm{n}(f)\,\d f &=& \frac{1}{\mu}\;\mathrm{n_0}\, \left ( \frac{f}{\mu}
\right) \;\frac{\d f}{\mu} \nonumber \\ 
&=& \mu^{-2}\,a_0\, \left( \frac{f}{\mu} \right)^{-s(f/\mu)}\, \d f
\end{eqnarray}
If $s$ does not vary appreciably over the interval $[f,f\mu]$, which is
well satisfied if $\mu$ departs only weakly from unity, then
\begin{eqnarray}
\mathrm{n}(f)\,\d f   &=& \mu^{s(f)-2}\,n_0(f)\,\d f\,.
\end{eqnarray}
Expressing this as a function of magnitude, we recover the form appropriate
for a magnitude limited sample (\citealt{1989ApJ...339L..53N}):
\begin{eqnarray}
\mathrm{N}(m)\,\d m  &=& \mu^{2.5\;s(m)-1}\,\mathrm{N_0}(m)\,\d m \nonumber \\
&=& \mu^{\alpha(m)-1}\,\mathrm{n_0}(m)\,\d m,
\label{eq:eq_mag}
\end{eqnarray}
The final form of the exponent ($\alpha(m) - 1$) reflects the two distinct 
effects of magnification and how they interact to produce the signal observed 
on the sky: the amplification effect that varies as a function of the quasar 
magnitude and the dilution effect that is a constant regardless of magnitude.  
The combination of these two effects is the magnification bias.

\begin{figure}[b]
\includegraphics[height=3in,width=3.25in]{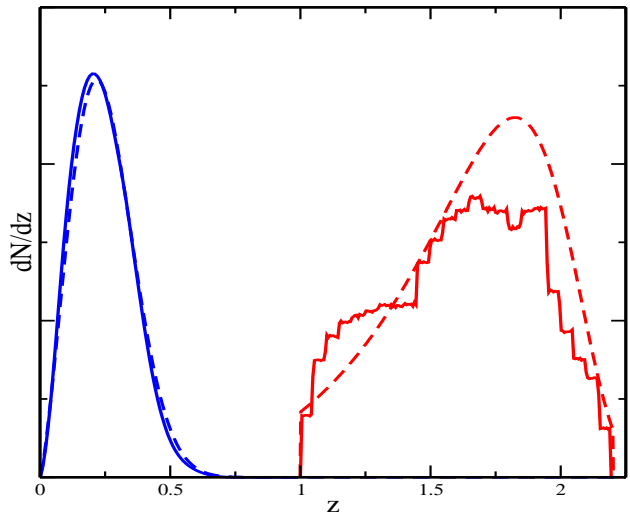}
\caption{Galaxy redshift distribution from applying our $17 < r < 21$ magnitude
limit to the CNOC2 luminosity function and quasar redshift distribution
inferred from quasar photometric redshifts (solid lines).  The fitted
redshift distributions from Equation~\ref{eq:dndz} are shown with dashed 
lines.  In all cases, the amplitude scaling is arbitrary.}
\label{fig:dNdzGalQSO}
\end{figure}

In the statistical context, magnification creates correlations between
foreground and background populations.  In the weak lensing regime,
i.e. if the convergence ($\kappa$) and the shear ($\gamma$) are
small compared to unity, a first-order Taylor expansion of the
magnification gives $\mu\approx1+2\,\kappa$.  Using this approximation,
a cross-correlation between magnification and foreground matter
overdensities can then be easily computed as a function of the matter
power spectrum.  The formalism for magnification by large-scale
structures was first introduced by \citet{bar95}.  Following the
prescription and notation laid out in \citet{jai03}, the
expected magnification bias signal can be written as
\begin{eqnarray}
w_{\rm GQ}(\theta,m) &=& 
12 \pi^2 \Omega_M \,(\alpha(m)-1) \times \nonumber \\
& & \int \d\chi~\d k~k\, {\cal K}(k, \theta, \chi)\, P_{gm}(k,\chi) \\
              &=& (\alpha(m)-1) \, \times w_0(\theta), \nonumber
\label{eq:theory_wgq}
\end{eqnarray}
where $m$ is the magnitude of the sources, $\Omega_M$ is the
cosmological matter density relative to critical, $\chi$ is the
comoving distance, ${\cal K}$ is the lensing kernel, and
$P_{gm}(k,\chi)$ is the galaxy-dark matter cross-power spectrum.  This 
formulation separates the expected signal into two pieces: $w_0(\theta)$,
which contains all of the information about non-linear galaxy biasing and
the redshift distributions; and $(\alpha(m) - 1)$, which varies only with
quasar magnitude and controls the sign of the expected signal.  We note in 
passing that $w_0$ is closely related to the tangential shear induced by the 
galaxy-mass correlation, which is measured in galaxy-galaxy lensing.  In 
practice, one needs to consider quasars over a given magnitude range, in 
which $\alpha(m)$ may vary. In this case we have:
\begin{eqnarray}
  w_{\rm GQ}(\theta) &=& 
\langle \alpha - 1 \rangle \times w_0(\theta),
\label{eq:theory_wgq2}
\end{eqnarray}
where
\begin{equation}
  \langle \alpha - 1 \rangle = \frac{\int \d m\, \mathrm{N}(m) 
    (\alpha(m) - 1)}
{\int \d m\, \mathrm{N}(m)}\,.
\label{eq:mean_alpha}
\end{equation}
The lensing kernel ${\cal K}$ in Eq. \ref{eq:theory_wgq} is primarily
a function of the redshift distributions of the galaxies and quasars.
As in \citet{jai03}, we model the redshift distributions as a
combination of power-law and exponential cut-off: $dN/dz \sim z^a \exp
(-(z/z_0)^b)$.  The galaxy redshift distribution can be inferred from
the luminosity functions measured by the CNOC2 survey (\citealt{lin99}) 
with the appropriate apparent magnitude limits (see \S\ref{sec:data}).  
For the quasars, photometric redshifts are computed for each quasar 
(see \S\ref{sec:data}), along with upper and lower redshift bounds and the 
probability that the quasar is within that redshift range.  To model the 
redshift distribution for each quasar sample, we assume a flat distribution
between the upper and lower redshift bounds for each quasar and weight
according to the aforementioned redshift probability.  To first order
the redshift distributions for all five quasar magnitude cuts are
indistinguishable, so we use the same fitted distribution to model
each cut, making $\langle \alpha - 1 \rangle$ the only free parameter
separating each magnitude bin.  Fitting the redshift distributions to
the chosen form, we find:
\begin{eqnarray}
\left (\frac{dN}{dz} \right)_G &{\sim}& 
z^{1.3} \exp \left [ -(z/0.26)^{2.17}\right ] \nonumber \\
\left ( \frac{dN}{dz} \right)_Q &{\sim}& 
z^{2.56} \exp \left [ -(z/2.02)^{12.76} \right ],
\label{eq:dndz}
\end{eqnarray}
where the redshift distribution for the quasars is limited to the range 
$1 < z < 2.2$ using photometric redshifts (see \S\ref{sec:data}).  
Figure~\ref{fig:dNdzGalQSO} shows the raw and fitted redshift distributions.

The only remaining piece of the theoretical calculation is the power spectrum.
Since we are in the non-linear regime of gravitational collapse for the 
smallest angular bins and our foreground redshift distribution is broad, a 
thorough calculation of the expected signal would involve a matter-galaxy 
power spectrum calculated with an evolving halo occupation distribution (HOD). 
However, for the purposes of this paper, we are only interested in checking 
our measurements against a simple model of the expected signal, leaving the 
task of extracting the proper HOD behavior for future papers.  With that in 
mind, we assume a simple HOD:
\begin{equation}
\left \langle N \right \rangle (M) = N_{c} + \left (M/M_0 \right )^\beta,
\end{equation}
where $N_{c}$ is unity for halo mass above $10^{11}h^{-1}
M_\sun$ and zero otherwise, $M_0$ is $10^{12}h^{-1} M_\sun$ and
$\beta$ is roughly unity.  These parameters are approximately what one
finds from semi-analytic galaxy codes (cf. \citealt{kra04})
and HOD fits to the SDSS spectroscopic survey 2-point clustering
measurements in \citet{zeh04}.

We will use this formalism and a flat WMAP cosmology ($\Omega_M = 0.29$, 
$\Omega_\Lambda = 0.71$, $h = 0.72$, $n = 1$;~\citealt{spe03}) to estimate
$w_0(\theta)$ and compare our measurements to the theoretical expectations. 
On small scales, non-linear magnification must be taken into account in order
to obtain a more accurate modeling (\citealt{men02b,tak03}), but this
level of precision will suffice for our current case.  A more detailed
modeling which includes marginalization over cosmological and redshift 
distribution parameters will be used in a future paper in order to constrain 
some of the model parameters.

With this model in hand, we can test whether the measured signal is
due to gravitational lensing in two ways: (i) we can test whether the
amplitude of the cross--correlation properly varies as a function of
magnitude, i.e. if $w_{\rm GQ}(m)\propto \langle \alpha(m)-1 \rangle$,
where $\langle \alpha(m) - 1 \rangle$ is evaluated according to
Eq.~\ref{eq:mean_alpha}; and (ii) we can check if the angular
variation of $w_{\rm GQ}(\theta)$ agrees with theoretical
expectations. Showing that the signal satisfies these two conditions
is a robust test to demonstrate the lensing origin of the signal and a
general lack of systematic contamination.

\section{Data Analysis}\label{sec:data}

The use of large, homogeneous samples of galaxies and quasars observed 
by the SDSS (\citealt{gun98, fuk96, smi02}) improves on previous 
measurements of the cosmic magnification in several key ways.  First, the 
SDSS provides accurate multi-color photometry (\citealt{lup99,ive04,hog01,
pie03}) over large areas of sky with tight control of the systematic 
errors that could alter the observed density of sources on the sky (e.g., 
seeing variations, masks around bright stars, sky background subtraction 
problems around bright galaxies).  Moreover, the accuracy of the photometry 
is crucial, especially to determine the number of faint sources.  If the 
required photometric accuracy is a few percent, CCD-based photometry
is clearly superior to photographic plate data. For example, the 2dF 
photometric accuracy is approximately 0.2 mag for objects with 
$17<\rm{b}_j<19.45$.  The corresponding incompleteness introduces an extra 
scatter for the density of sources on small scales and can thus mimic a 
signal.  

The second advantage of the SDSS comes from the multi-color photometry.  As
described below, consistent color-based selection over the full photometric
survey allows us to select a larger (both in quasar numbers and in area), more 
uniform quasar sample than has ever been compiled for cross-correlation 
studies.  This is critical both for minimizing Poisson errors on the 
measurement as well as avoiding systematic selection effects.  In addition, 
the multi-color photometry allows for the reliable estimation of photometric 
redshifts for quasars, removing any redshift overlap between our galaxy and 
quasar populations.  Given the fact that correlations due to intrinsic 
clustering have a much larger amplitude than the ones expected from lensing, 
even a small fraction of background sources at low redshift can give rise 
to a positive amplitude bias in the cross-correlation.  

\subsection{The Data}\label{sec:data_sub}

The data set was drawn from the third SDSS data release (DR3;~\citealt{aba03}).
Before masking, this set covers roughly 5000 square degrees, the majority of 
which is located around the North Galactic Cap.  To limit our contamination 
from systematic errors (\citealt{scr02}) in the photometric data, we imposed a 
seeing limit of 1\farcs4 and an extinction limit of 0.2 in the $r$ band.  We 
also included a mask blocking a $60''$ radius around bright galaxies 
($r < 16$) and stars with saturated centers to avoid losing quasars due to 
local fluctuations in sky brightness and observing defects (\citealt{man05}).  
The combination of these masks reduced our total area to $\sim 3800$ square 
degrees.

With these systematic cuts, we can reliably perform star/galaxy
separation using Baye\-sian methods to $r = 21$ (\citealt{scr02}). 
These selection criteria yielded 13.5 million galaxies between $17 < r < 21$ 
at a density of approximately one galaxy per square arcminute.  For galaxies,
we use {\tt counts\_model} magnitudes, while quasar magnitudes are given 
using {\tt psfcounts} magnitudes (\citealt{sto02}; these magnitudes are 
designated as {\it modelMag} and {\it psfMag} in the SDSS database, 
respectively).  In all cases, we de-redden the magnitudes to correct for 
Galactic dust extinction before applying the various magnitude cuts.  For 
even modestly faint magnitudes ($r > 18$), the Petrosian magnitudes used 
in the SDSS spectroscopic sample (and the previous SDSS galaxy-quasar 
measurements by \citealt{gaz03}) can fluctuate with the local seeing, 
leading to a strong variation ($\sim$ 25\%) in apparent galaxy density with 
seeing for a magnitude-limited sample.  The apertures used for 
{\tt counts\_model} magnitudes are convolved with the local PSF which makes 
them much more robust against seeing variations.  For the seeing range 
between 0\farcs85 and 1\farcs4, the observed galaxy density as a function 
of local seeing is constant for our magnitude cut.  As mentioned in 
\S\ref{sec:modeling}, applying these apparent magnitude cuts to the CNOC2 
luminosity functions yields a mean redshift for this magnitude limited sample 
of $z \sim 0.3$, with the maximum redshift of the sample near $z \sim 0.75$

The quasar data set was generated using kernel density estimation
(KDE) methods described in \citet{ric04}.  Although our quasar sample is drawn
from DR3, the selection method is identical to the one \citet{ric04} applied
to the DR1 data set.  The KDE method is a sophisticated extension of the 
traditional color selection technique for identifying quasars.  In this 
implementation, two training sets, one for stars and one for quasars are 
defined.  Then the colors for each new object are compared to those of each 
object in each training set and a 4D Euclidean distance is computed with 
respect to the objects in each training set.  New objects are then classified 
in a binary manner (quasar/star) according to which training set has a larger 
probability of membership.  This technique allowed clean separation of 
relatively low redshift ($z \leq 2.5$) quasars from the stellar locus in 4
dimensional color space, producing a catalog of 225,000 quasars down
to a limiting magnitude of 21 in the $g$ band with greater efficiency
and completeness than the SDSS spectroscopic targeting algorithm
(\citealt{ric02, bla03}).  After masking, the total population
was reduced to 195,000 quasars.

In addition to finding quasars, we applied photometric redshift
techniques (\citealt{wei04}) to filter out low redshift quasars which
might be physically associated with our foreground sample.  Given the
broader features and larger redshift range for quasars relative to
those of galaxies, the photometric redshift errors for the quasars are
generally somewhat asymmetric.  Rather than estimate a Gaussian
redshift error, we used an upper and lower redshift bound along with
the likelihood that the redshift was within those bounds.  To prevent
redshift overlap with the galaxies, we required that the upper and
lower bounds were within the range $1 < z < 2.2$ and weighted each
quasar according to the aforementioned redshift likelihood for both the 
number count and cross-correlation measurements.

\subsection{Measurement}\label{sec:measurement}

The expected lensing signal for magnification bias is generally
dominated on small scales ($ < 0.01^\circ$) by Poisson noise and falls
below the noise on scales larger than $1^\circ$.  To cover this full
range (and beyond), we used two estimators.  For angular scales below
$0.1^\circ$, we used a pair-based estimator similar to the
Landy-Szalay estimator (\citealt{lan93}), but modified for a
cross-correlation:
\begin{equation}
w_{\rm GQ}(\theta) = \frac{\langle GQ \rangle - \langle R_G Q \rangle - 
  \langle G R_Q \rangle + \langle R_G R_Q \rangle} {\langle R_G R_Q \rangle},
\label{eq:xcorr_small}
\end{equation}
where $\langle GQ \rangle$ is the number of galaxy-quasar pairs separated by 
angle $\theta$, $\langle R_G R_Q \rangle$ is the number of pairs of randomized
galaxy and quasar positions separated by $\theta$, etc.  To limit the Poisson
noise in our estimation of $\langle R_G Q \rangle$, $\langle G R_Q \rangle$
and $\langle R_G R_Q \rangle$, we generated 50 random points for each galaxy
and quasar.

As we move from small to large scales, the estimator in 
Equation~\ref{eq:xcorr_small} becomes progressively less and less efficient;
As the angular scale increases, a progressively larger area much be searched 
for suitable pairs, and the estimator in Equation~\ref{eq:xcorr_small} becomes
less and less efficient, increasing computation time.  Thus, for anglar bins 
larger than $0.05^\circ$, we used a pixel-based estimator.  Calculating the 
fractional galaxy and quasar over-densities ($\delta_G$ and $\delta_Q$, 
respectively), $w_{\rm GQ}$ is given by 
\begin{equation}
w_{\rm GQ}(\theta) = \frac{\sum_{i,j} \delta_{G,i} \delta_{Q,j} f_i f_j}{f_i f_j
},
\label{eq:xcorr_large}
\end{equation}
where we sum over all pairs of pixels separated by angle $\theta$ and 
$f_i$ is the fraction of pixel $i$ that remains after masking.  This angular
split roughly divides the total computation time for all of the various 
sub-samples (see \S\ref{sec:results}) equally between the large and small 
angle estimator codes.

For both estimators, we used 30 jack-knife samples (\citealt{scr02}) to 
generate errors, allowing us to combine the large and small angular 
measurements (including the single overlapping angular bin) to generate a 
coherent covariance matrix ($C(\theta_\alpha,\theta_\beta)$),
\begin{eqnarray}
C(\theta_\alpha,\theta_\beta) &=& \left (\frac{N}{N-1}\right )^2 \times 
\nonumber \\
&{\sum_{i=1}^N}&\left (w_{{\rm GQ},i}(\theta_\alpha) - 
\bar{w}_{GQ}(\theta_\alpha) \right) \times \nonumber \\
& & \left (w_{{\rm GQ},i}(\theta_\beta) - 
\bar{w}_{GQ}(\theta_\beta) \right),
\label{eq:jack}
\end{eqnarray}
where $N$ is the number of jack-knife samples and
$\bar{w}_{GQ}(\theta_\alpha)$ is the average value of 
$w_{{\rm GQ},i}(\theta)$ for all $N$ samples.  Equation~\ref{eq:jack} measures 
the variance directly on the sky, so it should capture the contribution from 
the cross-correlation as well as the galaxy and quasar auto-correlations.  
We expect the errors to be dominated by Poisson noise, with subdominant terms 
coming from cosmic variance as well as lensing by foreground structure not 
contained in our galaxy sample.  To increase our sensitivity at small angles 
where the signal is most interesting, we employed a hybrid logarithmic 
binning scheme.  For the angular decade running from $0.001^\circ$ to 
$0.01^\circ$, we used 3 logarithmically spaced bins, 4 bins for $0.01^\circ$ to
$0.1^\circ$, 5 bins for $0.1^\circ$ to $1^\circ$, etc.  This improved the 
signal-to-noise on small scales at the expense of generating a slightly 
larger off-diagonal elements in the covariance matrix than that produced 
by a straight logarithmic binning system.  However, the covariance matrices 
remained invertible in all cases with no degenerate modes, allowing us to 
use them for significance testing and curve fitting with no complications.

\begin{figure}
\includegraphics[height=3in,width=3.25in]{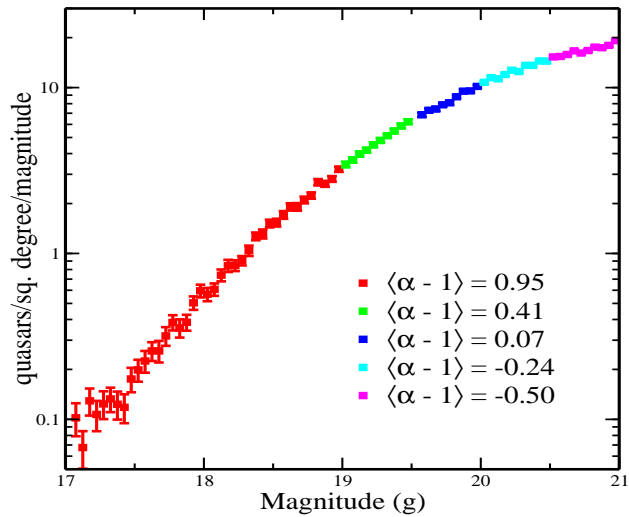}
\caption{Weighted number counts per magnitude per square degree in the 5 
magnitude bins for quasars with photometric redshift between $1 < z < 2.2$.}
\label{fig:NumberCountQSO_mag_bins}
\end{figure} 

\section{Results}\label{sec:results}
\subsection{Lensing Origin of the Signal}

In order to investigate the lensing origin of the signal, we first
measured the quasar-galaxy correlations $w_{\rm GQ}(\theta,m)$ as a
function of quasar magnitude in a given band. In Figure
\ref{fig:NumberCountQSO_mag_bins} we show the number counts of quasars
as a function of magnitude in the $g$ band.  We separated the $g$-selected 
quasar sample into five magnitude ranges and estimated the corresponding value 
of $\langle \alpha - 1 \rangle$ from power law fits in each bin.  The results 
are presented in Table~\ref{tab:g_counts}.  As can be seen, the values of 
$\langle \alpha - 1 \rangle$ are greater than zero for the three
brighter magnitude bins.  Therefore, due to the magnification bias, we
expect to find an excess of such quasars in the vicinity of foreground
lenses. In a similar way, we expect a deficit of quasars with
$g>20$. 

\begin{table}[h]
\begin{center}
\begin{tabular}{cc}
 \hline\hline
Magnitude & $\langle \alpha - 1 \rangle $\\\hline
$17~~<~g~<~19~~$ & $~0.95$\\
$19~~<~g~<~19.5$ & $~0.41$\\
$19.5<~g~<~20~~$ & $~0.07$\\
$20~~<~g~<~20.5$ & $-0.24$\\
$20.5<~g~<~21~~$ & $-0.50$\\
\hline \hline
\end{tabular}
\caption{The weighted mean value of $\alpha(m) - 1 = 2.5\,\mathrm{d log
    N_0}(m)/\mathrm{d}m - 1$ obtained from power law fits in different
  magnitude bins in the $g$ band. The values obtained in all five bands
  are presented in Table~\ref{tab:qso_counts}.}
\end{center}
\label{tab:g_counts}
\end{table}

\begin{figure}
\begin{center}
\includegraphics[height=3in,width=3.25in]{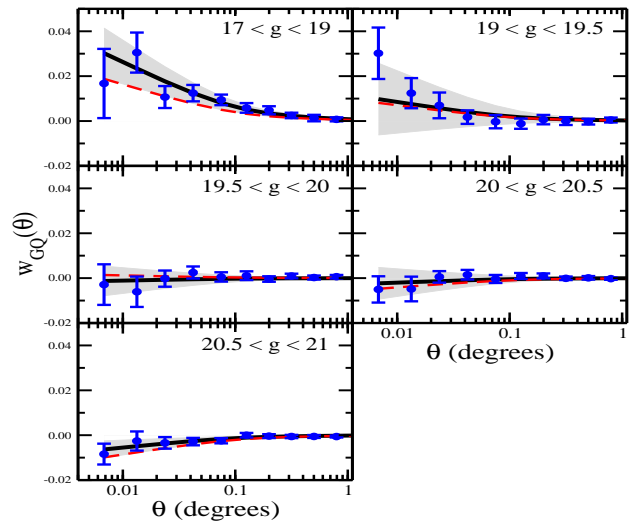}
\end{center}
\caption{Measurements of $w_{\rm GQ}(\theta)$ as a function of quasar $g$ band 
magnitude.  Error bars are the $1\sigma$ errors based on the jack-knife 
covariance (Equation~\ref{eq:jack}).  The dark solid curve is the fitting model
and the light dashed curve is the expected curve from the number counts
$\langle \alpha - 1 \rangle$.  The shaded region indicates the $1\sigma$ range 
on the fitted value of $\langle \alpha - 1 \rangle$.  Fitted and expected 
values for each magnitude bin are given in Table~\ref{tab:qso_counts}.  For 
angular scales larger than 1 degree, the measurements were consistent with 
zero in all five magnitude bins.}
\label{fig:GalQSO_fit}
\end{figure}

The corresponding quasar-galaxy correlation functions in each
magnitude bin are shown in Figure~\ref{fig:GalQSO_fit}.  As expected,
the brightest quasar sample with the steepest number count slope
showed the strongest positive cross-correlation, with the signal
amplitude dropping and eventually changing to an anti-correlation as
$\langle \alpha - 1 \rangle$ decreases.  At large angles, we see a
flat signal consistent with zero for all five quasar magnitude bins.
This first test verified {\it qualitatively} that the measured signal
satisfied the first of the two criteria described at the end of
\S\ref{sec:modeling}: amplitude variation as a function of quasar
magnitude.

\begin{figure}
\includegraphics[height=3in,width=3.25in]{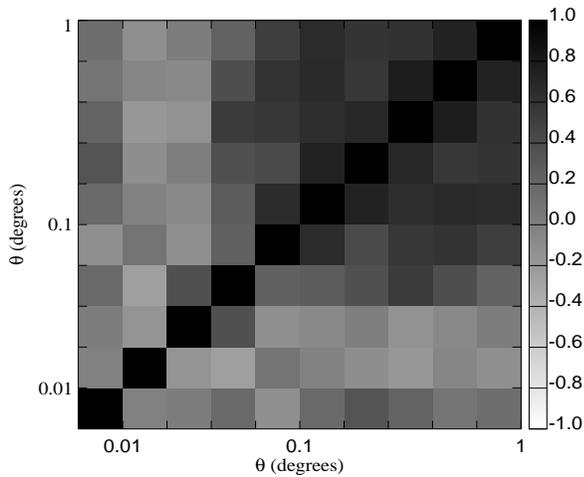}
\caption{Normalized covariance matrix for the $17 < g < 19$ measurement of
$w_{\rm GQ}(\theta)$.  For a given pair of angular bins ($i$ and $j$), the
normalized covariance matrix is given by 
$C(\theta_i,\theta_j)/\sqrt{C(\theta_i,\theta_i) C(\theta_j,\theta_j)}$.  The
level of correlation between angular bins is roughly consistent for all of
the measurements of $w_{\rm GQ}(\theta)$.}
\label{fig:CorrWthetaGalQSO}
\end{figure} 

We can now quantify this agreement by using the model given in
\S\ref{sec:modeling} and the covariance matrices measured using  
Equation~\ref{eq:jack} (see Figure~\ref{fig:CorrWthetaGalQSO}) to fit 
the measured data points and estimate the value of 
$\langle \alpha - 1 \rangle$ in each magnitude bin.  These
fits are shown with the solid black line in Figure~\ref{fig:GalQSO_fit} and 
the one-sigma uncertainty by the shaded region.  The value of the parameter 
$\langle \alpha - 1 \rangle$ obtained in this manner can be compared to the 
one directly measured from the quasar number counts.  The expected measurement
based on the quasar number counts is shown by the dashed red line.  For all 
five $g$-selected magnitude bins, we find agreement between the fitted and 
measured values of $\langle \alpha - 1 \rangle$ as a function of quasar
magnitude.  This demonstrates that the behavior of the signal quantitatively
follows both the amplitude and the angular variations expected from 
magnification bias.

\begin{figure}
\includegraphics[height=3in,width=3.25in]{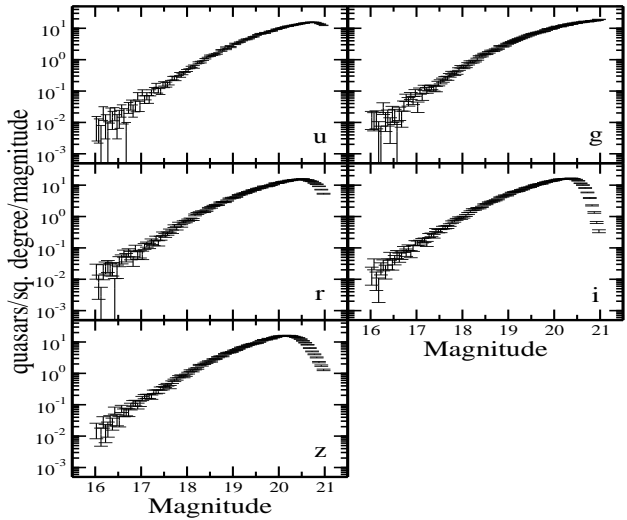}
\caption{Number counts per magnitude per square degree in the 5 filters
for quasars with photometric redshift between $1 < z < 2.2$.  The original
sample is magnitude limited in the $g$ filter.  The effective color cuts 
resulting from the quasar selection lead to incompleteness in the other 
filters at the faint end.}
\label{fig:NumberCountQSO_all}
\end{figure} 

We repeated similar measurements using magnitude-limited samples in each of 
the other four SDSS bands.  The quasar number counts in all five bands are 
given in Figure~\ref{fig:NumberCountQSO_all}.  As mentioned above, the quasars
were magnitude limited in $g$.  For the other bands, the combination
of effective color cuts for the sample and intrinsic scatter led to
strong incompleteness for magnitudes fainter than 20.  For the purpose
of separating the sample into magnitude bins, the turn-over point set
the faintest limit for each band.  The results for these measurements are
summarized in Table~\ref{tab:qso_counts}.  As with the $g$-selected 
measurements, cross-correlations in the other four filters found qualitative
and quantitative agreement with the expected magnitude bias variation with 
$\langle \alpha - 1 \rangle$.

\subsection{Optimal Stacking and Detection Significance}\label{sec:detection}

\begin{figure}
\includegraphics[height=3in,width=3.25in]{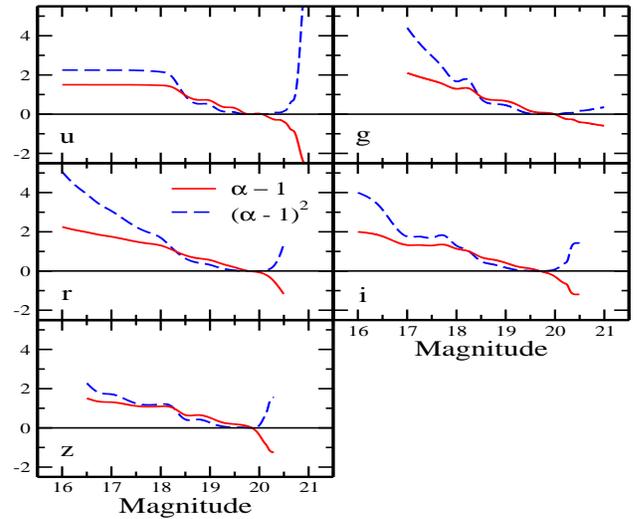}
\caption{$\alpha - 1$ and $(\alpha - 1)^2$ as a function of magnitude in the 5 
filters for quasars with photometric redshift between $1 < z < 2.2$.  For the
$u$, $r$, $i$, and $z$ bands, incompleteness in the number counts at the faint
end causes $\alpha - 1$ to rapidly diverge, making a direct measurement of the 
expected anti-correlation in these filters very difficult.}
\label{fig:NumberCountAlpha_all}
\end{figure}

Having shown that the signal follows the theoretical expectations 
as a function of magnitude, we combined these measurements to quantify the 
significance of the global detection of cosmic magnification.  Instead of 
separating the signal into five magnitude bins, we measured the signal 
integrated over all magnitudes weighted with different powers of 
$(\alpha(m)-1)^n$.  In Figure~\ref{fig:NumberCountQSO_all}, we show
the number count relations in each of the five SDSS filters and 
Figure~\ref{fig:NumberCountAlpha_all} plots the corresponding values of
$(\alpha(m)-1)^n$ for $n=1$ and $2$.  These plots are made by measuring 
$\langle \alpha - 1 \rangle$ in narrow magnitude bins over the full range and
then interpolating over the bins with a cubic spline.

\begin{figure}
\includegraphics[height=3in,width=3.25in]{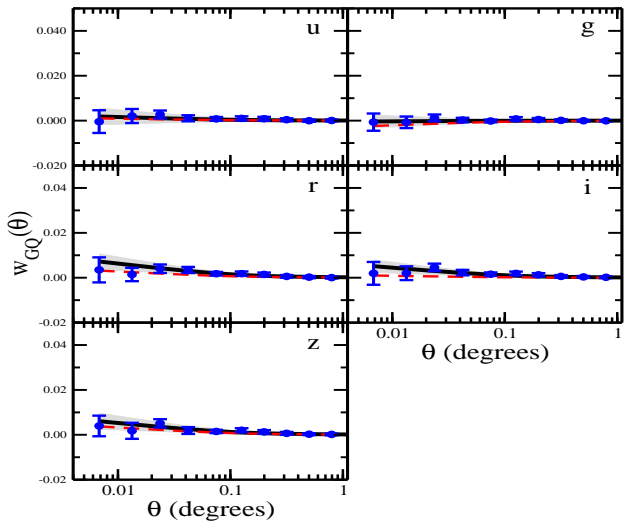}
\caption{Same as Figure~\ref{fig:GalQSO_fit}, but over the full magnitude 
range in each wavelength band.  The range on the y-axis is the same as in 
Figure~\ref{fig:GalQSO_fit} to provide an easier comparison of the relative
amplitude for samples where we do and do not expect a strong lensing signal.  
Fitted and expected values for $\langle \alpha - 1 \rangle$ are given in 
Table~\ref{tab:qso_counts}.  The correlation between angular bins for each 
measurement is consistent with that shown in Figure~\ref{fig:CorrWthetaGalQSO}
to first order.}
\label{fig:GalQSO_full}
\end{figure}

$\bullet$ Mean correlation function: by simply averaging the signal
from all quasars, i.e. considering the case $n=1$, we recover
Equation~\ref{eq:theory_wgq}, where $\langle \alpha - 1 \rangle$ is
given by integration over the full magnitude range of the sample.  The
effects of bright and faint quasars generally canceled each other,
resulting in the small values of $\langle \alpha - 1 \rangle$ found in
Table~\ref{tab:qso_counts}.  Figure~\ref{fig:GalQSO_full} shows the
results for all five SDSS filters, along with expected and fitted
curves for $\langle \alpha - 1 \rangle$.  As with the
magnitude-selected samples, we see generally good agreement between
the expected and observed signals.  Note that all the data points
are below the one-percent level.  

\begin{figure}
\includegraphics[height=3in,width=3.25in]{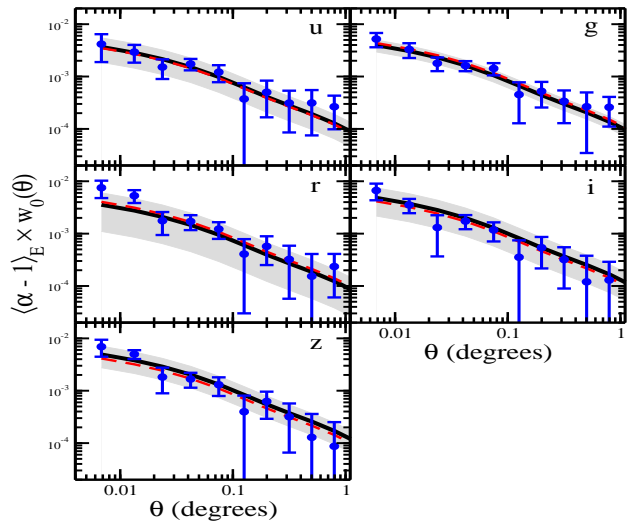}
\caption{Same as Figure~\ref{fig:GalQSO_full}, but for 
$w_{\rm GQ}^{\rm optimal}(\theta)$, which uses quadratic weighting to enhance
the lensing S/N (Equation~\ref{eq:galqso_norm}).  Fitted and expected values 
for $\langle \alpha - 1 \rangle_E$ are given in Table~\ref{tab:qso_counts}.  
The correlation between angular bins for each measurement is consistent with 
that shown in Figure~\ref{fig:CorrWthetaGalQSO} to first order.}
\label{fig:GalQSO_norm}
\end{figure}

$\bullet$ Optimal correlation function: as shown by \citet{men02a},
using $n=2$ (i.e. looking at the second-order moment of the signal as
a function of magnitude) optimally weights the expected lensing
signal.  This maximizes the S/N of the detection since the signal is
weighted proportionally to the expectations. 
With the extra factor of $\alpha(m) - 1$, the expected signal is:
\begin{eqnarray}
w_{\rm GQ}^{\rm optimal} (\theta) &=& 
\langle (\alpha(m)-1)^2\rangle
\times w_0(\theta)\nonumber\\ 
&=&\frac{\int \d m\, N(m)\,(\alpha(m)-1)^2}
{\int \d m\, N(m)}\times w_0(\theta)\nonumber\\
&=& \langle \alpha - 1 \rangle_E \times w_0(\theta)\,.
\label{eq:galqso_norm}
\end{eqnarray}
The corresponding signal can be measured by weighting each quasar by a
factor of $\alpha(m) - 1$ and by calculating the cross-correlation in the
manner described by Equations~\ref{eq:xcorr_small} and
\ref{eq:xcorr_large}.  Rather than largely counter-acting each other
as seen in Figure~\ref{fig:GalQSO_full}, the positive and negative
correlations from the bright and faint end of the quasar number counts
now act in concert and benefit from the statistical power of the
entire quasar population.  The corresponding results are presented in
Figure~\ref{fig:GalQSO_norm}.  Once again, we find a very good
agreement between the model and the observations for all five bands.

Using this optimally-weighted correlation function and the associated
covariance matrix, we can assess the significance of our detection. By
comparing the corresponding $\chi^2$ values of $w_{\rm GQ}^{\rm
optimal}(\theta)$ against the null for 18 angular bins, we detect the
signal at 4.1$\sigma$, 8.1$\sigma$, 4.8$\sigma$, 5.4$\sigma$ and
4.8$\sigma$ in the $u$, $g$, $r$, $i$ and $z$ bands respectively.
If we consider only the angular scales $\le 1^\circ$, as shown in 
Figure~\ref{fig:GalQSO_norm}, the significance of the detection remains 
nearly the same. 

Given the high S/N provided by the optimally-weighted estimator,
we can compare the angular variation of the measured signal to
theoretical expectations. As can be seen in Fig.\ref{fig:GalQSO_norm},
we find consistency from $0.3'$ to $1\deg$, i.e. over more than two
orders of magnitude in scale. This allows us to validate the second
criterion from \S\ref{sec:modeling}: the match of the angular
variation of the signal with the predicted cross-correlation function.
Considering an effective redshift of $z=0.3$ for the foreground galaxy
population, we find that the detected magnification signal probes scales
ranging from $\sim 60 h^{-1}$ kpc to $10$ Mpc.

Observing \emph{both} the first and second moments of the lensing
signal as a function of magnitude to give the expected behavior as a
function of the observed values of $\alpha(m)-1$ is an excellent
indication that we are observing the signal originating from
gravitational lensing.

\subsection{Systematics}

As seen in \S\ref{sec:intro}, accurate measurements of galaxy and
quasar number counts can suffer from a number of biases: seeing
variations, stellar contamination, dust extinction, redshift overlap,
etc.  To verify that our measurements are not affected by these effects, we
have performed a number of checks.  

To test for stellar contamination in our sample of quasars, we cross-correlated
stars in the $17 < r < 21$ range with the $g$ band selected quasars,
$w_{\rm SQ}(\theta)$, both with and without optimal weighting.  Unlike 
galaxies or quasars, the local stellar density is not well approximated by the
global mean density.  As a result, we do not expect (and do not
observe) a null correlation between stars and quasars (or stars and
galaxies).  Rather, our observed $w_{\rm SQ}(\theta)$ was consistent
with the observed galaxy-star cross-correlation, both of which are consistent
with a very small ($\sim$ 1\%) level of stellar contamination.  More 
importantly, when we optimally weighted $w_{\rm SQ}(\theta)$, the signal was
consistent with zero at all angular scales, as would be expected for a
stellar density independent of $\alpha(m) - 1$.  This was in marked
contrast to $w_{\rm GQ}^{\rm optimal}(\theta)$, detected at $8\sigma$.  
Cross-correlations with local seeing produced similar results.

We also tested the robustness of the photometric redshift likelihood by
applying a series of redshift likelihood thresholds (i.e. requiring that the
probability that the quasar was within the upper and lower redshift ranges
specified by the quasar photometric redshift algorithm was above a given 
value: 50\%, 60\%, 70\%, etc.).  In all cases, the resulting galaxy-quasar 
cross-correlation was consistent with measurements made with no threshold, 
verifying that our signal was not dominated by low probability outliers.  
Finally, a cross-correlation with low redshift quasars produced a large 
amplitude, positive signal.  However, this last point was sensitive to a 
restrictive cut on the quasar redshift probability ($> 80\%$) due to a 
strong shift in the probability distributions for quasars below 
$z \sim 1$; higher redshift quasars tended to have much higher redshift 
probabilities ($\sim 0.8$) than low redshift quasars (peaks around 0.5 and 
0.8).

Next, we checked for possible contamination by large, bright galaxies.
As described in \citet{man05}, the estimation of the density of faint
sources around bright extended objects can be biased induced by
uncertainties in the sky subtraction. We do not expect the sky subtraction 
issues to be as significant since our quasars are point sources, but, as 
described in \S\ref{sec:data}, we applied a 60$''$ mask around all $r < 16$ 
galaxies.  Measurements with and without these masks were identical, but we
included the masks in our final analysis to avoid any unforeseen effects.

Finally, a bias that is not related to the data analysis but that might be 
intrinsically present is extinction by dust. Indeed, the presence of dust 
around galaxies is expected to redden and extinct background sources. So far, 
the amount of dust on large scales has been poorly constrained and its 
effects on measurements of quasar-galaxy correlations has been uncertain. 
However, our analysis indicates that the deficit of quasars due to dust 
extinction is subdominant to the density changes induced by gravitational
lensing.  Indeed, the fact that the measured signal for the
\emph{first and second} moments behaves as expected as a function of the
slope number counts, $\alpha(m)-1$, indicates that the signal might
not be contaminated by other sources than gravitational lensing.
Biases like dust extinction or the above-mentioned effects are not
expected to scale proportionally to $\alpha(m)-1$ and would therefore
affect the first and second moments of the signal in different ways.
This would prevent the simultaneous agreements found above. Therefore
we conclude that our current measurements are not significantly affected 
by biases, but a parallel effort is underway to quantify the reddening effects
of the lensing galaxies more precisely.

\section{Discussion \& Future Applications} \label{sec:discussion}

In this paper we have presented a detection of cosmic magnification
obtained by cross-correlating distant quasars and foreground galaxies.
Using data from approximately 3800 square degrees of the SDSS
photometric sample, we have cross-correlated the position of
$\sim200,000$ photometrically selected quasars and large-scale
structures traced by over 13 million galaxies, and we have detected a
signal on angular scales from 20$''$ to 1 degree at high significance.

The magnification bias due to weak lensing gives rise to an excess or a 
deficit of background sources in the vicinity of foreground galaxies 
depending on the value of the power-law slope of the source number counts:
$\mathrm{\alpha}(m)=2.5\,\mathrm{d log\, N_0}(m)/\mathrm{d}m$.  Our
measurements of the galaxy-quasar cross-correlation function exhibit
the expected behavior: bright quasars, with steep number counts,
appear to be in excess around galaxies and large-scale structures, and
faint quasars with shallow number counts are seen to be in deficit. On
all scales, we find $w_{\rm GQ}(m) \propto (\alpha(m)-1)$ in the five SDSS
bands, as expected.

We have measured the first and second moments of the signal as a
function of quasar magnitude and the results are in very good
agreement with what is expected from the magnification bias: depending
on the band, the first moment gives an amplitude consistent with zero
or smaller than $\sim 5 \times 10^{-3}$, as a result of the opposite effects
arising from the bright and the faint quasars. The second moment,
which turns out to be the optimal signal estimator, exhibits a strong
signal detected at $>4\sigma$ in all five SDSS filters and reaching up
to 8.1$\sigma$ in the $g$-band. The quasars are magnitude selected in the 
$g$-band, giving us the largest sample in this band (other bands lose quasars
at the faint end due to the effective color cuts in these bands), so the 
difference in the signal-to-noise in the other bands relative to $g$ is 
unsurprising.  Using this estimator we find the angular dependence of the 
signal to be in very good agreement with theoretical estimations of lensing 
by large-scale structures. Our measurements probe physical scales ranging 
from $60\,h^{-1}\,$kpc to $10\,h^{-1}\,$Mpc at the mean lens redshift.  Since
we do not expect the biases from systematic errors to scale proportionally
to $\alpha(m)-1$, the simultaneous agreement of the first and second moments 
of the signal as a function of magnitude indicates that these systematic biases
(including dust extinction) do not significantly affect our measurements.

The SDSS quasar and galaxy samples used in our analysis are significantly
larger, more uniform and better characterised than any data sets used for
this measurement previously. We have shown that biases including seeing
variations, stellar contamination, sky subtraction issues and errors in
the photometric redshifts are well controled and do not significantly
affect the measurements. Whereas previously claimed detections reported a
signal much larger than theoretical predictions, our measurement shows,
for the first time, the expected amplitude and angular dependence for the 
standard cosmological model and a realistic galaxy biasing. As such, we 
conclude that the disagreement between theoretical predictions and previous 
measurements was most likely due to larger systematic effects in these data 
sets which could not be adequately controled.

The successful detection of cosmic magnification opens the door to a
number of applications: as mentioned in \S\ref{sec:modeling}, cosmic 
magnification is a function of the first moment of the galaxy halo occupation
distribution (HOD) on all angular scales.  Conversely, the galaxy
auto-correlation function, $w(\theta)$, is a strong function of the
second moment of the HOD.  Thus, measuring these two quantities for
the same sample of galaxies will provide us with constrains on both
moments, and therefore probe the scales on which the galaxy biasing
becomes stochastic.  Such an analysis can then be carried out as a
function of galaxy type, redshift, etc. and provide interesting
constraints on our understanding of galaxies and large-scale
structures.

As noted in \S\ref{sec:intro}, our measurements of cosmic magnification
constrain the projected galaxy-mass correlation in much the same way
as galaxy-galaxy lensing, although that method is based on galaxy
shapes and measurements of shear.  This complementarity is a
particularly useful cross-check since the dominant sources of
systematic error for the two methods are different (PSF anisotropy
for the galaxy-galaxy shear measurements vs. photometric calibration
for the magnification bias).  Furthermore, using quasars as sources,
cosmic magnification allows for probing lensing at higher redshifts:
the \emph{sources} used for SDSS galaxy-galaxy lensing studies are
used as \emph{lenses} for measurements of quasar-galaxy correlations.
Finally, as is the case for cosmic shear, higher-order statistics can
also be investigated in the context of lensing-induced quasar-galaxy
correlations (\citealt{men03}). We also note that the techniques used
for efficient quasar selection are readily applicable to next
generation of large, multi-band surveys. Cosmic magnification is
therefore an excellent complement to planned cosmic shear surveys.

In a future work we will use measurements of quasar-galaxy
correlations to generate the first constraints on the galaxy HOD and 
cosmological parameters obtained from cosmic magnification.  Likewise, 
projects are underway to use cosmic magnification to measure the extent of 
galaxy dust halos as well dark matter halo ellipticities.

\section*{Acknowledgments}

The authors would like to thank Daniel Eisenstein, Michael Jarvis, 
Rachel Mandelbaum, Yannick Mellier and Michael Strauss for useful comments.

RS and AJC acknowledge partial support from the NFS through CAREER award 
AST9984924 and ITR grant 1120201.

BM acknowledges the Florence Gould foundation for its financial 
support

ADM and RJB acknowledge support from NASA through grants NAG5-12578 and
NAG5-12580 and through the NSF PACI Project.

Funding for the creation and distribution of the SDSS Archive has been provided
by the Alfred P. Sloan Foundation, the Participating Institutions, the National
Aeronautics and Space Administration, the National Science Foundation, the U.S.
Department of Energy, the Japanese Monbukagakusho, and the Max Planck Society. 
The SDSS Web site is http://www.sdss.org/.

The SDSS is managed by the Astrophysical Research Consortium (ARC) for the 
Participating Institutions.  The Participating Institutions are The University 
of Chicago, Fermilab, the Institute for Advanced Study, the Japan Participation
Group, The Johns Hopkins University, the Korean Scientist Group, Los Alamos 
National Laboratory, the Max-Planck-Institute for Astronomy (MPIA), the 
Max-Planck-Institute for Astrophysics (MPA), New Mexico State University, 
University of Pittsburgh,  University of Portsmouth,  Princeton University, 
the United States Naval Observatory, and the University of Washington.

\tabletypesize{\normalsize}
\begin{deluxetable*}{ccccc}
%\tabletypesize{\scriptsize} 
\tablecolumns{5} 
%\tablewidth{0pt}
\tablecaption{Quasar counts, $\langle \alpha - 1 \rangle$ from the number 
count slopes in each magnitude bin, $\langle \alpha - 1 \rangle$ from fitting 
observed $w_{\rm GQ}(\theta)$ to the weak-lensing model given in 
Equation~\ref{eq:theory_wgq2} and detection significance for the optimal
estimator in each filter.  For the optimal estimator, the full magnitude range 
is used and the values of $\langle \alpha - 1 \rangle$ are for 
$\langle \alpha - 1 \rangle_E$ as given in Equation~\ref{eq:galqso_norm}.  
Detection significance is given in multiples of standard deviation ($\sigma$).}
\startdata
\hline\hline\\[1mm]
Magnitude Limit & Quasar Counts & $\langle \alpha - 1 \rangle$ & 
Fitted $\langle \alpha - 1 \rangle$ & $\sigma$\\[1mm]\hline\\[1mm]
$17 < u < 19$   & 6774 & $+0.97$ & $+1.63 \pm 0.70$ &  \\[1mm]
$19 < u < 19.5$ & 9001 & $+0.42$ & $+0.19 \pm 0.95$ &  \\[1mm]
$19.5 < u < 20$ & 16648 & $+0.07$ & $-0.27 \pm 0.39$ &  \\[1mm]
$20 < u < 20.3$ & 14824 & $-0.11$ & $+0.04 \pm 0.55$ &  \\[1mm]
$20.3 < u < 20.6$ & 18524 & $-0.37$ & $-0.08 \pm 0.34$ &  \\[1mm]
$17 < u < 20.6$   & 65610 & $+0.05$ & $+0.09 \pm 0.19$ &  \\[1mm]\hline\\[1mm]
Optimal              & 65610 & $+0.18$ & $+0.19 \pm 0.09$ & {\bf 4.1} \\[1mm]
\hline \hline\\[1mm]
$17 < g < 19$   & 8054 & $+0.95$ & $+1.53 \pm 0.57$ &  \\[1mm]
$19 < g < 19.5$ & 10312 & $+0.41$ & $+0.49 \pm 0.81$ &  \\[1mm]
$19.5 < g < 20$ & 18148 & $+0.07$ & $-0.06 \pm 0.33$ &  \\[1mm]
$20 < g < 20.5$ & 28751 & $-0.24$ & $-0.12 \pm 0.36$ &  \\[1mm]
$20.5 < g < 21$ & 39567 & $-0.50$ & $-0.32 \pm 0.19$ &  \\[1mm]
$17 < g < 21$   & 104683 & $-0.12$ & $-0.02 \pm 0.14$ &  \\[1mm] \hline\\[1mm]
Optimal              & 104683 & $+0.22$ & $+0.20 \pm 0.05$ & {\bf 8.1} \\[1mm] 
\hline \hline\\[1mm]
$16 < r < 18.5$ & 4212 & $+1.19$ & $+1.84 \pm 0.79$ &  \\[1mm]
$18.5 < r < 19$ & 6120 & $+0.65$ & $+1.36 \pm 0.86$ &  \\[1mm]
$19 < r < 19.5$ & 12101 & $+0.35$ & $-0.30 \pm 0.60$ &  \\[1mm]
$19.5 < r < 20$ & 21141 & $+0.03$ & $+0.14 \pm 0.25$ &  \\[1mm]
$20 < r < 20.3$ & 18137 & $-0.25$ & $-0.11 \pm 0.34$ &  \\[1mm]
$16 < r < 20.3$ & 61596 & $+0.16$ & $+0.37 \pm 0.17$ &  \\[1mm] \hline\\[1mm]
Optimal              & 61596 & $+0.20$ & $+0.18 \pm 0.12$ & {\bf 4.8} \\[1mm]
\hline \hline\\[1mm]
$16 < i < 18.5$ & 5609 & $+1.08$ & $+1.66 \pm 0.81$ & \\[1mm]
$18.5 < i < 19$ & 7813 & $+0.56$ & $+1.06 \pm 0.96$ &  \\[1mm]
$19 < i < 19.5$ & 15236 & $+0.23$ & $-0.10 \pm 0.76$ &  \\[1mm]
$19.5 < i < 20$ & 26173 & $-0.06$ & $+0.21 \pm 0.33$ &  \\[1mm]
$20 < i < 20.2$ & 17687 & $-0.51$ & $-0.11 \pm 0.57$ &  \\[1mm]
$16 < i < 20.2$ & 72391 & $+0.05$ & $+0.26 \pm 0.14$ & \\[1mm] \hline\\[1mm]
Optimal              & 72391 & $+0.21$ & $+0.24 \pm 0.13$ & {\bf 5.3} \\[1mm]
\hline \hline\\[1mm]
$16 < z < 18.5$ & 5812 & $+1.00$ & $+1.58 \pm 0.77$ &  \\[1mm]
$18.5 < z < 19$ & 8047 & $+0.62$ & $+0.97 \pm 0.85$ &  \\[1mm]
$19 < z < 19.5$ & 16056 & $+0.29$ & $-0.05 \pm 0.37$ &  \\[1mm]
$19.5 < z < 19.8$ & 15240 & $+0.13$ & $-0.02 \pm 0.39$ & \\[1mm]
$19.8 < z < 20.1$ & 20177 & $-0.27$ & $-0.25 \pm 0.49$ &  \\[1mm]
$16 < z < 20.1$ & 65207 & $+0.19$ & $+0.31 \pm 0.18$ &  \\[1mm] \hline\\[1mm]
Optimal              & 65207 & $+0.21$ & $+0.25 \pm 0.11$ & {\bf 4.8} \\[1mm]
\hline\hline
\enddata 
\label{tab:qso_counts}
\end{deluxetable*}

\end{document}